\documentclass[conference]{IEEEtran}
\IEEEoverridecommandlockouts

\usepackage{cite}
\usepackage[utf8]{inputenc}
\usepackage{amsmath,amssymb,amsfonts}
\usepackage{pifont}
\usepackage{booktabs}
\usepackage{array}
\usepackage{caption}
\usepackage{algorithmic}
\usepackage{graphicx}
\usepackage{mathrsfs}
\usepackage{textcomp}
\usepackage[most]{tcolorbox}
\usepackage{xcolor}
\usepackage{subfig}
\definecolor{customGreen}{HTML}{5cb85c}
\definecolor{customRed}{HTML}{ed4337}
\usepackage{setspace}
\usepackage{hyperref}
\usepackage{comment}
\usepackage{setspace}

\usepackage{fancyhdr}
\usepackage[norelsize, linesnumbered, ruled, lined, boxed, commentsnumbered, noend]{algorithm2e}
\definecolor{customGreen}{HTML}{00FF00}
\definecolor{customRed}{HTML}{FFE4E1}
\definecolor{customYellow}{HTML}{FFF2CC}
\definecolor{customViolet}{HTML}{E1D5E7}
\definecolor{newGreen}{HTML}{D5E8D4}
\definecolor{newRed}{HTML}{F8CECC}

\newcommand{\cmark}{\ding{51}}  
\newcommand{\xmark}{\ding{55}}  

\usepackage[font=footnotesize,labelfont=bf,skip=0pt]{caption}
\begin{document}

\pagenumbering{arabic}
\title{Minimizing Energy in Reliability and Deadline-Ensured Workflow Scheduling in Cloud}

\author{\IEEEauthorblockN{Suvarthi Sarkar, Dhanesh V,  Ketan Singh, Aryabartta Sahu \textit{IEEE Senior Member}}
\IEEEauthorblockA{\textit{Dept. of CSE,  IIT Guwahati, Assam, India.} Emails:\{s.sarkar, dhanesh.v, s.ketan, asahu\}@iitg.ac.in }
}

\maketitle
\setstretch{0.9}
\thispagestyle{plain}
\pagestyle{plain}

\begin{abstract}
With the increasing prevalence of computationally intensive workflows in cloud environments, it has become crucial for cloud platforms to optimize energy consumption while ensuring the feasibility of user workflow schedules with respect to strict deadlines and reliability constraints. The key challenges faced when cloud systems provide virtual machines of varying levels of reliability, energy consumption, processing frequencies, and computing capabilities to execute tasks of these workflows. To address these issues, we propose an adaptive strategy based on maximum fan-out ratio considering the slack of tasks and deadline distribution for scheduling workflows in a single cloud platform, intending to minimise energy consumption while ensuring strict reliability and deadline constraints. We also propose an approach for dynamic scheduling of workflow using the rolling horizon concept to consider the dynamic execution time of tasks of the workflow where the actual task execution time at run time is shorter than worst-case execution time in most of the cases. Our proposed static approach outperforms the state-of-the-art (SOTA) by up to 70\% on average in scenarios without deadline constraints, and achieves an improvement of approximately 2\% in deadline-constrained cases. The dynamic variant of our approach demonstrates even stronger performance, surpassing SOTA by 82\% in non-deadline scenarios and by up to 27\% on average when deadline constraints are enforced. Furthermore, in comparison with the static optimal solution, our static approach yields results within a factor of 1.1, while the dynamic approach surpasses the optimal baseline by an average of 25\%. 


\end{abstract}


\section{Introduction}

Cloud computing has become increasingly popular, enabling users to offload applications to remote servers in exchange for a subscription fee. This model benefits both users by reducing execution costs and cloud providers by ensuring a steady revenue stream. However, with rising demand, cloud platforms face significant challenges in managing user workloads efficiently. Failure to deliver reliable and timely services can result in users migrating to competing providers.

Many user applications are represented as Directed Acyclic Graphs (DAGs), commonly referred to as workflows, which express task-level dependencies. These workflows are typically deadline-constrained and reliability-critical—any failure incurs substantial penalties, and guaranteed reliability is essential. Furthermore, workflows are often computationally intensive and difficult to manage.

Although cloud providers offer virtually unlimited computing capacity by deploying Virtual Machines (VMs) on Physical Machines (PMs), VMs are still vulnerable to hardware failures, network disruptions, and system crashes, thereby affecting overall service reliability. Moreover, energy consumption constitutes nearly 50\% of the total operational cost for cloud providers \cite{Power_model}, making energy efficiency a critical concern. Reducing energy consumption while maintaining reliability has become a key optimization objective for modern cloud service providers. Dynamic Voltage and Frequency Scaling (DVFS) is the most used technique to reduce energy consumption. DVFS  dynamically adjusts the voltage and frequency of the system to match the workload, reducing power consumption when  running a light workload, but 
DVFS can potentially impact reliability, as lowering the voltage and frequency might increase the risk of errors or instability, especially in complex systems \cite{sota_rel}.

A growing number of users now employ cloud platforms to execute scientific workflows, which are inherently data-dependent \cite{trial}. This uncertainty in execution time adds further complexity, making traditional static scheduling approaches less effective.

In this work, we address the objectives of minimizing energy consumption and ensuring workflow reliability within deadline constraints in workflows with data-dependent execution times. By exploiting execution time variability and employing a rolling-horizon strategy \cite{Roll-H}, our solution dynamically adjusts resource allocation to minimize energy without compromising reliability. 

The following are the contributions of this work:
\begin{itemize}
    \item We formulate the energy minimization problem for deadline and reliability-constrained workflows in cloud systems. To the best of our knowledge, this problem is not addressed in the existing literature.
    
    \item We propose to explore the task replication strategy for reliability using primary and backup approaches \cite{pbmodel} considering the frequency scaling of the VMs.
    \item  We analyze how energy consumption is influenced by workflow fan-out patterns and we propose an adaptive strategy based on the maximum fan-out ratio considering slack of tasks and deadline distribution for scheduling workflows.
   \item We also propose an approach for dynamic scheduling of workflow using the rolling horizon concept to consider the dynamic execution time of tasks of the workflow where the actual task execution time at run time is shorter than worst-case execution time in most of the cases.
    \item Our proposed approach significantly outperforms existing methods when evaluated with workflow benchmark datasets. In scenarios without deadline constraints, our static approach surpasses the state-of-the-art (SOTA) by up to 70\%, and improves by about 2\% in deadline-constrained cases. The dynamic variant shows even stronger results, exceeding SOTA by 82\% in non-deadline scenarios and by up to 27\% under deadline constraints. Compared to the static optimal solution, our static approach is within 1.1 times, while the dynamic approach outperforms the static optimal baseline by an average of 25\%.
\end{itemize}

\begin{table}[tb!]
\caption{Summary of Literature Review (DL: Deadline, Rel.: Reliability, Dyn.: Dynamic) \label{tab:lit-summary}}
\centering
\footnotesize
\tabcolsep1.7pt 
\begin{tabular}{|p{2.2cm}|p{2.7cm}|c|c|c|c|}
\hline
\textbf{Paper} & \textbf{Method} & \textbf{DL} & \textbf{Rel.} & \textbf{Energy} & \textbf{Dyn.} \\

\hline
Cao \textit{et al.} \cite{ECao} & Genetic Algorithm  & \xmark & \cmark & \cmark & \xmark \\
\hline
Lingjuan \textit{et al.} \cite{sota_rel} & Reliability Distribution & \xmark & \cmark & \cmark & \xmark \\
\hline
Niar \textit{et al.} \cite{sota_deadline} & Deadline Distribution & \cmark & \xmark & \cmark & \xmark \\
\hline
Niyati \textit{et al.} \cite{Niyati} & Binary Search & \cmark & \cmark & \xmark & \xmark \\
\hline
Swain \textit{et al.} \cite{swain} & Efficient Replication & \cmark & \cmark & \xmark & \xmark \\
\hline
Mousavi \textit{et al.}\cite{sota_deadline_Mousavi} & Critical Path & \cmark & \cmark & \xmark & \xmark \\
\hline
Xiaojin \textit{et al.} \cite{Realtime} & Feedback Mechanism & \cmark & \xmark & \cmark & \cmark \\
\hline
\end{tabular}
\end{table}
\section{Literature Review} \label{sec:litreview}
Workflow scheduling under energy, reliability, and deadline constraints has become a key research focus in cloud computing, with numerous approaches proposed in recent literature. Below, we briefly discuss a few representative works in this area.

\nocite{pc6,pc7,pc8,pc9}
\subsection{Energy and Reliability Optimised Scheduling}
Zhu \textit{et al.} \cite{Roll-H} proposed an energy-efficient rolling horizon scheduling strategy for data centres. Cao \textit{et al.} \cite{ECao} introduced a genetic algorithm-based method to minimise workflow execution cost on VMs by adjusting CPU frequencies, balancing energy efficiency with reliability, and particularly addressing soft error risks due to reduced voltages. Lingjuan \textit{et al.} \cite{sota_rel} proposed a DVFS and frequency-scaling-based reliability distribution approach for energy minimisation under reliability constraints, though it omits deadline considerations, limiting practical applicability. Niar \textit{et al.} \cite{sota_deadline} addressed deadline and reliability-constrained multi-workflow scheduling using frequency scaling and VM mapping, aiming to reduce energy and rental costs through a composite score function. However, its focus on both user and provider costs complicates adoption.

Ghose \textit{et al.} \cite{Niyati} proposed a binary search-based strategy to determine the optimal number of replicas for meeting reliability constraints, considering MTTF and soft error models, while leaving dynamic scheduling as future work. Swain \textit{et al.} \cite{swain} focused on efficient task-to-PM mapping for workflows under varying machine failure rates. Mousavi \textit{et al.} \cite{sota_deadline_Mousavi} and Medara \textit{et al.} \cite{sota_deadline_Medara} developed critical-path-based techniques leveraging idle resources and DVFS, respectively, to improve reliability and minimise energy usage for single workflows.

\subsection{Real-Time Scheduling}
Real-time scheduling approaches offer dynamic adaptability in cloud environments. Xiaojin \textit{et al.} \cite{Realtime} proposed a feedback-driven algorithm for scheduling multiple workflows in real time under deadline constraints, optimising rental costs. Liu \textit{et al.} \cite{On_Line} introduced a non-preemptive utility accrual scheduling method using profit and penalty Time Utility Functions (TUFs) to dynamically evaluate task execution decisions.

A brief summary of the literature review is given in \autoref{tab:lit-summary}. While existing works explore various combinations of energy, reliability, and deadlines, none address energy minimisation for data-dependent workflows under reliability and deadline constraints in a real-time setting. This motivates the focus of our proposed approach.

\section{System Model and Problem Statement}
\begin{figure}
    \centering    \includegraphics[scale=0.9]{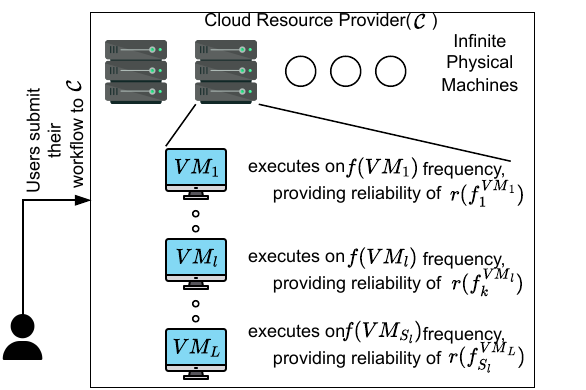}
    \caption{Users submit workflows to a cloud resource provider $\mathcal{C}$. $\mathcal{C}$ has access to an unlimited pool of Physical Machines (PMs), each capable of hosting virtual machines (VMs) of any type from a predefined set of $L$ VM types. 
}
    \label{fig:system-model}
\end{figure}
\subsection{System Model}
\label{sec:system_model}

In practical scenarios, cloud service providers such as AWS, Azure, and GCP are equipped with extensive computational resources, often approximated as an infinite pool of resources. We consider a cloud service provider denoted as $C$, possessing a theoretically unlimited number of physical machines (PMs). These PMs host various virtual machines (VMs) that are leased to users for executing computational tasks. We assume there are $L$ types of VMs available from the cloud service provider is represented formally as $\text{VM}(\mathcal{C}) = \{{ \text{VM}_1, \text{VM}_2, \dots, \text{VM}_l, \dots, \text{VM}_L }\}$, where $\text{VM}_l$ denotes the $l^{th}$ type of VM. Typically, cloud service providers classify VMs into categories such as tiny, small, medium, large, extra large depending on their computational capabilities. Also, we consider VMs with reliability ranges low, medium, high, very high, etc. The considered $L$ types of VM is Cartesian product of VM categorizes with compute power and  VM categorizes with different reliablity types.

Each VM type $\text{VM}_l$ can operate at a discrete set of frequency levels (scaled between 0 and 1), denoted by $f(\text{VM}_l) = \{{f_1^{\text{VM}_l}, f_2^{\text{VM}_l}, \dots, f_k^{\text{VM}l}\}, \dots, f_{S_l}^{\text{VM}_l}}$, as detailed in prior studies \cite{VM-Model,sota_deadline,sota_rel}. We consider that there are $S_l$ numbers of discrete frequency levels. Additionally, every VM type $\text{VM}_l$ has an associated compute power denoted as $\text{CP}_l$, provided explicitly by the cloud service provider. Each VM instance is restricted to executing only one task at a given moment. \autoref{fig:system-model} shows the proposed system architecture.

For simplicity and without loss of generality, we assume negligible data-transfer times among VMs due to the presence of high-speed interconnects within a single cloud environment. Furthermore, we presume no constraint on the number of VM instances that can be provisioned; thus, any number of VM instances of a specified type and frequency level can be provisioned from cloud service provider $\mathcal{C}$ \cite{Inf_PM,InfPM2}, effectively simulating an environment of unlimited computational resources.

\subsection{Task Model}
\label{sec:task_model}

We consider a workflow $W$ composed of a set of $N$ tasks $\mathbf{T} = \{{t_1, t_2, \dots, t_j, \dots, t_{N}}\}$, where $t_j$ represents the $j^{th}$ task of the $W$. The workflow $W$ is characterized by three parameters: arrival time $A_w$, required reliability $R_w$, and deadline $D_w$. The workflow can also be represented as a directed acyclic graph (DAG) $W = (\mathbf{T}, \mathbf{E})$, where $\mathbf{T}$ represents tasks and $\mathbf{E} = { e_{r,s}: r, s = 1, \dots, N }$ represents dependency edges connecting tasks. For any two tasks $(t_r, t_s)$ in $\mathbf{T}$, the execution order must respect the predecessor-successor relationship. An edge $e_{r,s}$ indicates that task $t_s$ can start only after task $t_r$ completes and the associated data transfer $DM_{r,s}$ from $t_r$ to $t_s$ occurs. We denote the predecessors of task $t_j$ as $\textit{pred}(t_j)$ and successors as $\textit{succ}(t_j)$. As stated previously, we assume negligible data-transfer time within a single cloud environment.

In alignment with real-world scenarios, we assume each task $t_j$ has a data-dependent worst-case execution length $wc_j$ measured in MI (Million Instructions). This worst-case length follows an unknown probability distribution $P_j$, influenced by factors such as input data characteristics. Consequently, task execution length is bounded by $wc_j$ but may conclude sooner with some positive probability depending on the actual data input \cite{wcCitation}.
 
\subsection{Energy Model}
\label{sec:energy_model}

The power consumption of a VM type $\text{VM}_l$ while executing a task $t_j$ can be computed according to the power model given in \cite{Power_model}. The total power consumed comprises two distinct components: the static power, which is the baseline power consumption during VM initialization, and the dynamic power, which varies based on computational workload. Formally, the power consumption per unit time for $\text{VM}_l$ is expressed as:

\begin{equation} \label{eq:power_model}
P(f_k^{\text{VM}_l}) = \alpha_l + \beta_l (f_k^{\text{VM}_l})^3
\end{equation}

where $\alpha_l$ denotes the static power component, and $\beta_l (f_k^{\text{VM}_l})^3$ represents the dynamic power component.

The total energy consumption associated with executing a task $t_j$ on $\text{VM}_l$ is defined by:

\begin{equation} \label{eq:energy_model}
E(t_j, f_k^{\text{VM}_l}) = P(f_k^{\text{VM}_l}) \cdot \tau(t_j, f_k^{\text{VM}_l})
\end{equation}

Here, the execution time $\tau(t_j, f_k^{\text{VM}_l})$ for task $t_j$, characterized by its data-dependent computational length $wc_j$ (in MI), when operating at frequency $f_k^{\text{VM}_l}$ on a $\text{VM}_l$ with computational capability $\text{CP}_l$, is given by: $\tau(t_j, f_k^{\text{VM}_l}) = \frac{wc_j}{\text{CP}_l \cdot f_k^{\text{VM}_l}}$.

\subsection{Reliability Model}
Reliability is a critical parameter in assessing the performance and dependability of cloud service providers. Reliability is influenced primarily by the hardware conditions of PMs, including factors such as temperature variations and power failures \cite{TransFail-Cite,ReliabilityCite}. Additionally, reliability depends significantly on the operating frequency of VMs hosting tasks on the respective PMs. We adopt the reliability modelling approach described in previous literature \cite{ReliabilityCite,Niyati}. For a VM type $\text{VM}_l$ operating at frequency $f_k^{\text{VM}_l}$, the failure rate $r(f_k^{\text{VM}_l})$ is defined as follows:

\begin{equation}
\label{eq:fr_vm}
r(f_k^{\text{VM}_l}) = r_0^l \cdot 10^{\psi \cdot \frac{f_{\text{max}}^{\text{VM}_l} - f_k^{\text{VM}_l}}{f_{\text{max}}^{\text{VM}_l} - f_{\text{min}}^{\text{VM}_l}}}
\end{equation}

Here, $f_{\text{max}}^{\text{VM}_l}$ and $f_{\text{min}}^{\text{VM}_l}$ denote the maximum and minimum operational frequencies available for the VM type $\text{VM}_l$ from the frequency set $f(\text{VM}_l)$. The failure rate $r_0^l$ is obtained empirically by operating the VM at its maximum frequency and is typically provided by the cloud platform based on historical data. The hardware reliability coefficient $\psi$ is dependent on the PMs, due to factors its processor, memory, hard-disk, network, usage frequency etc. Consequently, the failure rate for $\text{VM}_l$ is a function of both the VM characteristics and the PM hardware on which it is hosted. When the scheduler chooses a VM with reliability parameters $\psi$ and $r_0^l$, the VM to PM resource management layer handles internally providing the specified configuration of VM to the scheduler. 
By the term task reliability ($R_j(f_k^{\text{VM}_l})$), we refer to reliability of a task $t_j$ executing at frequency $f_k$ on ${\text{VM}_l}$. It follows exponential distribution as mentioned in \cite{TransFail-Cite,ReliabilityCite} is expressed as:
\begin{equation}
\label{eq:task_rel}
R_j(f_k^{\text{VM}_l}) = \exp \left\{ -r(f_k^{\text{VM}_l}) \cdot \tau(t_j, f_k^{\text{VM}_l}) \right\}
\end{equation}

In this equation, $\tau(t_j, f_k^{\text{VM}_l})$ denotes the actual data-independent worst time execution time of task $t_j$ at frequency $f_k^{\text{VM}_l}$ on $\text{VM}_l$. VMs operating at higher frequencies generally exhibit increased reliability but at the cost of higher power consumption \cite{ReliabilityCite,Niyati}. 

To enhance reliability further, we consider a task replication mechanism following the primary-backup model, allowing each task to be replicated at most once, as additional replications increase energy consumption and resource usage significantly \cite{PB_Cite,Niyati}. Thus, the reliability of a task $t_j$ replicated at the same frequency is given by:
\begin{equation}
\label{eq:task_rel_rep}
R_j^{\text{rep}}(f_k^{\text{VM}_l}) = 1 - (1 - R_j)^2
\end{equation}
This formulation implies that task failure occurs only if both replicas fail simultaneously. However, replicating a task essentially doubles the associated energy consumption.

Hence, the overall workflow reliability $R(W)$ is computed as the product of individual task reliabilities:
\begin{equation}
\label{eq:workflow_reliability}
R(W) = \prod_{t_j \in W} R_j^{\text{eff}}
\end{equation}

Here, $R_j^{\text{eff}}$ equals $R_j$ if task $t_j$ is not replicated, or $R_j^{\text{rep}}$ if replication is performed. 

\subsection{Problem Statement}
\label{sec:problem_statement}
We formulate the problem of energy minimization for workflow scheduling under deadline and reliability constraints, considering data-dependent execution times. Energy minimization contributes to the maximization of profit, which is the major motive for all cloud owners. As the operating frequency of a VM increases, task execution time decreases, which leads to higher power consumption and improved reliability. Conversely, reducing the frequency lowers both energy consumption and reliability. This introduces a trade-off between energy efficiency and reliability. The optimization problem is formally defined as follows:
\begin{equation} \label{eq:reliable_workflow_scheduling}
\min E(W) = \sum_{j=1}^{N} \sum_{l=1}^{L} \sum_{k=1}^{S_l} x_{j,l,k} E(t_j,f_k^{\text{VM}_l}),
\end{equation}

subject to

   \begin{equation} \label{eq:workflow_reliability_condition}
   \footnotesize
   R(W) \geq R_w
   \end{equation}

   \begin{equation} \label{eq:start_finish_time_constraint}
   \footnotesize
   St(t_s) \geq \max{ (Ft(t_r))},  \quad e_{r,s} \in \mathbf{E}
   \end{equation}

   \begin{equation}
   \label{eq:dag-const}
   \footnotesize
       Ft(t_j) = St(t_j) + \sum_{l=1}^{K} \sum_{k=1}^{S_l} x_{j,l,k} \cdot \tau(t_j, f_k^{\text{VM}_l})
   \end{equation}

   \begin{equation} \label{eq:deadline_constraint}
   \footnotesize
   \max_{j}{Ft(t_j)} \leq D_w 
   \end{equation}

    \begin{equation} \label{eq:Arrival_constraint}
    \footnotesize
   \min_{j}{St(t_j)} \geq A_w 
   \end{equation}

   \begin{equation} \label{eq:scheduling_constraint}
   \footnotesize
    \sum_{l=1}^{K} \sum_{k=1}^{S_l} x_{j,l,k} = 1, \quad j = 1, \dots, N.
   \end{equation}

   \begin{equation*} \label{eq:scheduling_decision_variable}
   \footnotesize
   x_{j,l,k} = 
   \begin{cases} 
   1, & \begin{array}{l}
         \text{if } t_j \text{ is scheduled on } \text{VM}_l \\
          \text{ with frequency } f_k^{\text{VM}_l};
       \end{array} \\[10pt]
   0, & \text{otherwise}.
   \end{cases}
   \end{equation*}

The constraint defined by \autoref{eq:workflow_reliability_condition} imposes the hard reliability requirement. \autoref{eq:start_finish_time_constraint} enforces task execution order based on workflow data dependency edges. Moreover, \autoref{eq:deadline_constraint} guarantees that the workflow meets its deadline, \autoref{eq:Arrival_constraint} ensures no task begins execution before the workflow arrives, and constraint \autoref{eq:scheduling_constraint} mandates scheduling each task exactly once.

\nocite{pc1,pc2,pc3,pc4,pc5}
\section{Solution Approach}
\label{sec:sol-approach}

Before presenting our proposed solution approach, we introduce some terms which we used in out approach. These terms are instrumental in defining the temporal constraints of tasks in a workflow. Let \( \operatorname{VM}_\text{bst} \) denote the virtual machine with the highest compute power. All the terms are calculated using the worst-case execution time of the tasks.

\begin{itemize}
    \item Earliest Start Time (EST): The minimum time at which task \( t_j \) can begin execution. If \( t_j \) has no predecessors, then \( \operatorname{EST}(t_j) = A_w \), the workflow's arrival time. Otherwise,
    \begin{equation}
        \label{eq:est}
        \operatorname{EST}(t_j) = \max_{t_k \in \operatorname{pred}(t_j)} \operatorname{EFT}(t_k)
    \end{equation}
    
    \item Earliest Finish Time (EFT): The earliest possible completion time of task \( t_j \), given by:
    \begin{equation}
        \label{eq:eft}
        \operatorname{EFT}(t_j) = \operatorname{EST}(t_j) + \tau(t_j, f_\text{max}^{\operatorname{VM}_\text{bst}})
    \end{equation}
    where $VM_\text{bst}$ is the VM with the highest compute power.

        \item Latest Start Time (LST): The latest time task \( t_j \) can commence without causing delays in workflow completion:
    \begin{equation}
        \label{eq:lst}
        \operatorname{LST}(t_j) = \operatorname{LFT}(t_j) - \tau(t_j, f_\text{max}^{\operatorname{VM}_\text{bst}})
    \end{equation}
    
    \item Latest Finish Time (LFT): The maximum allowable time by which task \( t_j \) must finish without violating the workflow deadline. If \( t_j \) has no successors, then \( \operatorname{LFT}(t_j) = D_w \). Otherwise,
    \begin{equation}
        \label{eq:lft}
        \operatorname{LFT}(t_j) = \min_{t_k \in \operatorname{succ}(t_j)} \operatorname{LST}(t_k)
    \end{equation}
    
    The terms EST, LST, EFT and, LFT are as defined by the work of We et al.~\cite{LSTEFT}.
    
    \item Schedule (\( \mathcal{S} \)), Start Time (\( St(t_j) \)), Finish Time (\( Ft(t_j) \)): We define the schedule of a task \( t_j \) as \( \mathcal{S}(t_j) \gets \left( f_{k}^{\text{VM}_{l}}, St(t_j), Ft(t_j) \right) \), indicating that task \( t_j \) is assigned to virtual machine \( \text{VM}_l \), operating at frequency level \( f_{k}^{\text{VM}_l} \), with a start time of \( St(t_j) \) and a finish time of \( Ft(t_j) \).

\end{itemize}

\subsection{Overall Framework for Proposed Approach}
 Upon submission of a workflow $W$ by the user, the cloud service provider ($\mathcal{C}$) initiates the feasibility check using the method described in \autoref{subsec:bcpwsa}. If the workflow $W$ is deemed feasible, $\mathcal{C}$ computes the Earliest Start Time (EST), Earliest Finish Time (EFT), Latest Start Time (LST), and Latest Finish Time (LFT) using equations \autoref{eq:est}, \ref{eq:eft}, \ref{eq:lst}, and \ref{eq:lft}, respectively.

 $\mathcal{C}$ invokes the Adaptive Strategy Based on Maximum Fan-Out Ratio (ASMFR), as outlined in \autoref{alg:adaptive-mfr}, to determine which scheduling strategy to use between the two proposed methods: Largest Energy First (LEF) (shown in \autoref{alg:LEF-SLWSA}) and Level-based Deadline Distribution (LDD) (shown in \autoref{alg:DD-SLWSFA}). The decision is based on the parameter Maximum Fan-Out Ratio (MFR).

Using the selected strategy, $\mathcal{C}$ generates a task schedule that includes both primary and backup task-to-VM mappings. This schedule is then passed to a \textit{Rolling Horizon} framework \cite{Roll-H}, which manages the execution of all undispatched tasks. When a task becomes ready (i.e., all its predecessors have completed), it is dispatched to an appropriate VM hosted on a physical machine (PM) under the control of $\mathcal{C}$.

As each task completes execution—potentially before its worst-case execution time—$\mathcal{C}$ updates its actual finish time, observed successful task execution, and energy consumption. This updated information is then used to recompute the schedule for the remaining undispatched tasks using the same selected strategy. The newly updated schedule is passed back to the \textit{Rolling Horizon}, which continues dispatching the next set of ready tasks and monitoring task readiness dynamically.

\begin{figure}[tb!]
    \centering
    \includegraphics[width=\linewidth]{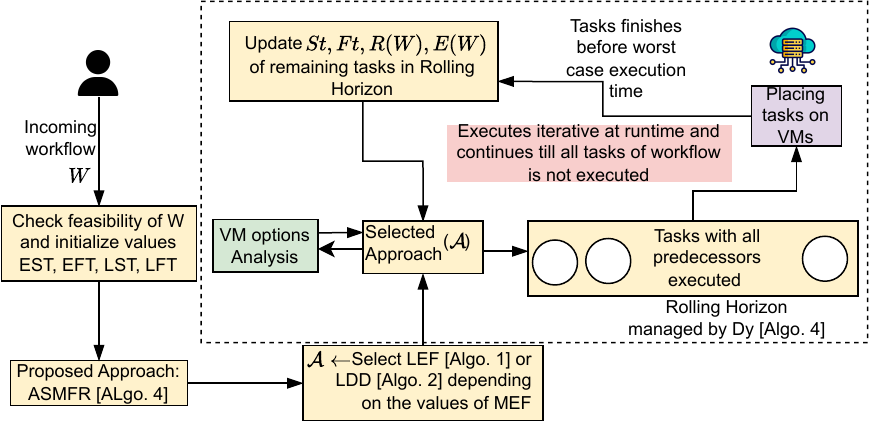}
\caption{Central Scheduler Architecture. The components enclosed within the dotted line represent dynamic events that occur at runtime and are written within \textcolor{newRed}{\rule{1.5ex}{1.5ex}} colour. System-related information is indicated using \textcolor{newGreen}{\rule{1.5ex}{1.5ex}} colour. Steps handled by our proposed approach are marked with \textcolor{customYellow}{\rule{1.5ex}{1.5ex}}, while operations managed by the cloud platform are denoted by \textcolor{customViolet}{\rule{1.5ex}{1.5ex}}.}

    \label{fig:scheduler-architecture}
\end{figure}

This overall framework enables real-time adaptability and optimization of energy usage while ensuring that both the deadline and reliability constraints of the workflow are upheld. The complete architectural overview is shown in \autoref{fig:scheduler-architecture}.

\subsection{Feasibility Check of incoming workflow}
\label{subsec:bcpwsa}

To evaluate the feasibility of a workflow \( W \), we begin scheduling each task on the virtual machine (VM) with the highest compute power at the highest operating frequency, we term this as BCP (Best Compute Capacity) schedule. As a result, all tasks commence execution at their earliest start time (EST) and complete at their earliest finish time (EFT), as defined in \autoref{eq:est} and \autoref{eq:eft}. If the workflow violates the deadline constraint under this schedule, it is deemed infeasible and subsequently rejected, since this baseline configuration already yields the minimum possible makespan. Additionally, we assess whether the workflow satisfies the required reliability constraint by allowing at most one task replication for each of the tasks, in line with our problem formulation. If both constraints are satisfied, $W$ is considered feasible, and we aim to minimise energy consumption in the subsequent sub-sections.

\subsection{Largest Energy First Approach (LEF)}
\label{subsec:lef}
In this approach, incorporating energy minimization strategies for feasible workflow $W$. We introduce the notion of \emph{slack} for a task $t_j$, denoted by $\mathscr{S}(t_j)$, which is defined as the difference between its latest finish time and the actual finish time obtained under the BCP schedule, i.e., $\mathscr{S}(t_j) = \text{LFT}(t_j) - Ft(t_j)$.

If $\mathscr{S}(t_j) > 0$, it indicates that there is an opportunity to reduce energy consumption by prolonging the execution of the task up to its LFT. To achieve this, we define a metric called \texttt{MIN-CPF}$(t_j)$, which stands for the minimum product of compute power and frequency (CP and $f$ as mentioned in \autoref{sec:system_model}) required to fully utilize the available slack without violating the deadline. It is computed as follows:
\begin{align}
\footnotesize
\label{eq:cpf}
\mathscr{S}(t_j) = \text{LFT}(t_j) - Ft(t_j) = 0 &\Rightarrow Ft(t_j) = \text{LFT}(t_j) \notag \\
\hspace*{-1.5em}\Rightarrow \frac{wc_j}{\texttt{MIN-CPF}(t_j)} &= \operatorname{LFT}(t_j) - St(t_j) \notag \\
\Rightarrow \texttt{MIN-CPF}(t_j) &= \frac{wc_j}{\operatorname{LFT}(t_j) - St(t_j)}
\end{align}

We try to execute tasks on the $\text{VM}_l$ operating near its, denoted by $f_\text{cri}^{\text{VM}_l}$. This critical frequency minimizes the energy usage for executing a task, as derived from \autoref{eq:energy_model}, and is defined by:
\begin{equation}
\label{eq:critical_f}
    f_\text{cri}^{\text{VM}_l} = \sqrt[3]{\frac{\alpha_l}{2\beta_l}}
\end{equation}
Frequencies on either side of $f_\text{cri}^{\text{VM}_l}$ result in higher energy consumption. However, due to the discrete nature of supported frequencies for each $\text{VM}_l$, it may not always be feasible to exactly match the required \texttt{MIN-CPF} or the critical frequency. In such cases, we select a frequency $f_k^{\text{VM}l}$ from the available set such that it is as close as possible to $f\text{cri}^{\text{VM}_l}$ while ensuring that $\text{CP}_l \cdot f_k^{\text{VM}_l} \geq \texttt{MIN-CPF}(t_j)$.

We now describe the Largest Energy First (LEF) approach, as outlined in \autoref{alg:LEF-SLWSA}. The method begins by generating an initial schedule $\mathcal{S}$ using the BCP approach to verify the feasibility of the workflow $W$.
Subsequently, it orders the tasks in $W$ in non-increasing order of their energy consumption based on its worst-case execution time and stores this ordered list in $\mathcal{L}$. This prioritization ensures that tasks with the highest energy consumption are considered first, allowing slack-based optimizations to yield maximal reductions in overall energy usage.

Additionally, we define an option list $\mathcal{O}$, which stores all the  feasible configurations (i.e., VM types and their corresponding feasible operating frequencies) for executing a given task. For each task in $\mathcal{L}$, the approach evaluates these options and selects the one that minimizes energy consumption while satisfying the deadline and reliability constraints. The tuple $(E^{\text{eff}}(t_j, f_{\textnormal{curr}}^{\text{VM}_{\textnormal{curr}}}) \textnormal{No Backup/ With Backup})$ signifies that task $t_j$ mapped to host on $VM_{curr}$ with operating frequency of $curr$, with or without backup depending of value of the second parameter. 

The approach iteratively applies this process to each task in $\mathcal{L}$. Our approach executes by the following steps listed below:

\begin{enumerate}

   \item Energy-aware scheduling near critical frequency:  We first aim to reduce energy consumption by executing tasks at or near their critical frequency. The minimum required to compute power-frequency product, denoted as \verb|MIN-CPF|$(t_j)$, is computed as per \autoref{eq:cpf}. For each VM type $\text{VM}_l$ available in the cloud, we identify the closest supported frequency $f_k^{\text{VM}_l}$ (using binary search) such that $\text{CP}_l \cdot f_k^{\text{VM}_l} \geq \texttt{MIN-CPF}(t_j)$.  As VM type and frequency selection influence both reliability and energy consumption (as mentioned in \autoref{eq:fr_vm}, \autoref{eq:task_rel}, and \autoref{eq:energy_model}), and reducing frequency decreases both power usage and reliability, extra care must be taken to meet the hard reliability constraint.
    
\item Ensuring reliability constraint: \label{it:rel-rep-step} We next verify whether the reliability constraint $R_w$ of workflow $W$ is satisfied. If not, we explore the option of replicating task $t_j$ at the same frequency as its primary, ensuring DAG constraints are respected (as in \autoref{eq:start_finish_time_constraint}).  In certain cases, replicating an earlier task ($t_{prev}$) in the list $\mathcal{L}$ may be more effective than replicating $t_j$ itself. Tasks towards the start of the list typically offer greater flexibility than tasks towards the other end, as the overall deadline is fixed. Thus, prioritizing the replication of such tasks can be more energy efficient as there is more chance that they execute in machines near the critical frequency.  The updated workflow reliability after modifying $t_j$’s VM-frequency mapping is computed by: 
\begin{equation}
\label{eq:rem_rel}
\footnotesize
    R_{\text{new}}(W) = 
    \left( \frac{R_{\text{old}}(W)}{R_j^{eff}(f_\text{old}^{\text{VM}_\text{old}})} \right) \cdot 
    R_j^{eff}(f_\text{new}^{\text{VM}_\text{new}})
\end{equation}
where $R_{old}(W)$ is the initial reliability of workflow $W$ when task $t_j$ was executing in $VM_{\text{old}}$ with frequency $f_\text{old}$ (i.e $R_j^{eff}(f_\text{old}^{\text{VM}_\text{old}})$), and  $R_{new}(W)$ is the modified reliability of $W$ when task $t_j$ is executing in $VM_{\text{new}}$ with frequency $f_\text{new}$ (i.e $R_j^{eff}(f_\text{new}^{\text{VM}_\text{new}})$).

 \item Effective energy consumption:  If a task is replicated, we must also account for the energy consumed by the replica. The effective energy consumption $E^{\text{eff}}(t_j, f_k^{\text{VM}_l})$ is calculated as:

\begin{equation}
\label{eq:eff_energy}
    =
    \begin{cases} 
        E(t_j,f_k^{\text{VM}_l}), \:\:\:\:\:\:\:\:\:\: \text{if $t_j$ is not replicated} \\
        2 \cdot E(t_j,f_k^{\text{VM}_l}), \:\: \:\:\:\text{if $t_j$ is replicated} \\
        E(t_j, f_\text{new}^{\text{VM}_\text{new}}) + E(t_\text{prev}, f_\text{prev}^{\text{VM}_\text{prev}}), \\ \:\:\:\:\:\:\:\:\:\:\:\:\:\:\:\:\:\:\:\:\:\:\: \text{if $t_{prev}$  is replicated}
    \end{cases}
\end{equation}
where  $t_\text{prev}$ is a previous task than $t_j$ in $\mathcal{L}$ with least energy $t_\text{prev}$ is replicated, and $f_\text{prev}^{\text{VM}_\text{prev}}$ is task $t_\text{prev}$'s frequency and VM mapping. For instance, consider a virtual machine \( \text{VM}_l \) with a maximum compute power (CP) of 5 MIPS and power parameters \( \alpha = 25 \) and \( \beta = 100 \), resulting in a critical frequency of 0.5 (as defined in \autoref{eq:critical_f}). Let tasks \( t_{j_1} \) and \( t_{j_2} \) have worst-case execution times of 20 and 10 MIPS, respectively. Executing \( t_{j_1} \) at the critical frequency of 0.5 results in an execution time of 4 time units and consumes 150 units of energy, calculated using \autoref{eq:energy_model} as:
$
E(t_{j_1}) = 4 \cdot (25 + 100 \cdot 0.5^3)
$
In contrast, executing \( t_{j_2} \) at frequency 1 yields an execution time of 2 time units but results in higher energy consumption of 350 units:
$
E(t_{j_2}) = 2 \cdot (25 + 100 \cdot 1^3)
$
This example illustrates that, despite a higher worst-case execution time, \( t_{j_1} \) consumes less energy due to being executed at a lower frequency. Hence, in certain scenarios, replicating tasks with larger execution times may be more energy-efficient.

\item Update start time and finish time of replicated task, its successors and predecessors:  The finish time of task $t_j$ is updated based on the selected VM and frequency: $Ft(t_j) = St(t_j) + \tau(t_j, \text{VM}_\text{new})$. The rescheduling of $t_j$ may affect the timing of its dependent tasks. Specifically, $St(.)$, EST, and EFT of its successors may increase, while LFT and LST of its predecessors may decrease. We propagate these updates using Breadth-First Search (BFS) on the workflow DAG.
\end{enumerate}

After this process, if the reliability constraint is still not satisfied, we replicate the tasks with minimum energy consumption until the constraint is met and return the final schedule $\mathcal{S}$.

\begin{algorithm} [tb!]
\setcounter{AlgoLine}{0}
\caption{Reliability and Energy Ensured Task Replication}
\footnotesize
\label{alg:ret}
\SetKwInOut{Input}{Input}
\SetKwInOut{Output}{Output}
\SetKw{KwInit}{Initialize}
\DontPrintSemicolon
\Input{$W = (\mathbf{T}, \mathbf{E})$, \( \text{VM}(C)\), EST, EFT, LST, LFT, MIN-CPF ($t_j$)}

\Output{Schedule of task $t_j$ \( \mathcal{S}(t_j) = (f_k^{\text{VM}_l}, St(t_j), Ft(t_j))\)}

        Find frequency $f_k^{\text{VM}_l}$ closer to $f_\text{cri}^{\text{VM}_l}$ (using \autoref{eq:critical_f}) such that $ \text{CP}_l \times f_\text{cri}^{\text{VM}_l} \geq \texttt{MIN-CPF}(t_j)$

        Calculate the updated frequency $R_\text{new}(W)$ using \autoref{eq:rem_rel}

        \KwInit{\textnormal{List} $\mathcal{H} \gets \emptyset$, \textnormal{$\mathcal{H}$ stores non-replicated scheduled tasks in ascending order of worst-case execution time}}

        \KwInit{\textnormal{Options list} $\mathcal{O} \gets \{ (E^{\text{eff}}(t_j, f_{\textnormal{curr}}^{\text{VM}_{\textnormal{curr}}}), \textnormal{No Backup}) \}$}
        
        \If{$R_{\text{new}}(W) \geq R_w$ \texttt{(when reliability is satisfied)}}{
            $\mathcal{O} \gets \mathcal{O} \cup \left\{ \left( E^{\text{eff}}(t_j, f_k^{\text{VM}_l}), \text{No Backup} \right) \right\}$
        }
        
        \Else{
            \If{\textnormal{Replicate $t_j$ with frequency level} $k$ \textnormal{such that} $R_{\text{new}}(W) \leq R_w$}{
                $\mathcal{O} \gets \mathcal{O} \cup \left\{ \left( E^{\text{eff}}(t_j, f_k^{\text{VM}_l}), \text{With Backup} \right) \right\}$
            }
            
            \If{\textnormal{Replicate} $t_\textnormal{prev}$ \textnormal{in} $\mathcal{H}$ \textnormal{with frequency level} $k$ \textnormal{such that} $R_{\text{new}}(W) \leq R_w$}{
                $\mathcal{O} \gets \mathcal{O} \cup \left\{ \left( E^{\text{eff}}(t_j, f_k^{\text{VM}_l}), \text{With Backup} \right) \right\}$
            }
        
    }
    
    Select $(E^{\text{eff}}(.),  \textnormal{Backup Option})$ \text{such that} $E^{\text{eff}}(.) = \displaystyle \min_{E}{\mathcal{O}}$ and reliability condition is maintained
    
    \If{\textnormal{no task is scheduled to run with backup copy}}{
        Insert $t_j$ in $\mathcal{H}$}
    \ElseIf{$t_\text{prev}$ \textnormal{is scheduled to run with backup copy}}
    {
    Remove $t_\text{prev}$ out of $\mathcal{H}$\\
    Insert $t_j$ in $\mathcal{H}$
    }

    Update $Ft(t_i), E(t_j, .), R_j^{\textnormal{eff}}, R(W)$ and $E(W)$ of $t_j$ based on the new mapping
    
    Change EST, $St(.)$ of successors and LFT, LST of ancestors of $t_j$

\Return \( \mathcal{S}(t_j) \)

\end{algorithm}

\begin{algorithm} [tb!]
\setcounter{AlgoLine}{0} 
\caption{Largest Energy First Approach (LEF)}
\label{alg:LEF-SLWSA}
\SetKwInOut{Input}{Input}
\SetKwInOut{Output}{Output}
\SetKw{KwInit}{Initialize}
\DontPrintSemicolon

\Input{$W = (\mathbf{T}, \mathbf{E})$, \( \text{VM}(C)\)}

\Output{Schedule \( \mathcal{S} \) for workflow \( W \) with \( \mathcal{S}(t_j) = (f_k^{\text{VM}_l}, St(t_j), Ft(t_j)), \: \forall t_j \in \mathbf{T}\)}

Calculate EST, EFT, LST, LFT of every task in $W$ using \autoref{eq:est}, \ref{eq:eft}, \ref{eq:lft}, \ref{eq:lst}.

\KwInit{$\mathcal{L} \gets$ \textnormal{Task in non-increasing order of their worst case execution time}}

\For{\textnormal{each task} \( t_j \) \textnormal{in list} $\mathcal{L}$}{

Calculate MIN-CPF ($t_j$) using \autoref{eq:cpf}

$\mathcal{S}(t_j) \gets$ Generate energy aware schedule for task $t_j$ ensuring deadline, reliability using \autoref{alg:ret} ($W$, \( \text{VM}(C)\), EST, EFT, LST, LFT, MIN-CPF ($t_j$))
}

\Return \( \mathcal{S}\)

\end{algorithm}

 \subsection{ Level-based Deadline Distribution Approach (LDD)}
 In the LEF approach, we aim to minimize the slack with respect to the latest finish time, $\operatorname{LFT}(t_j)$. Consider a bottleneck task $t_b$ (shown in \autoref{fig:wf} subfigure C) that has many successors and is selected from the list $\mathcal{L}$ for scheduling. Suppose we allocate it to a virtual machine $\text{VM}_b$ such that its slack $\mathscr{S}(t_b)$ becomes exactly zero. In this scenario, the finishing time of $t_b$ aligns precisely with its $\operatorname{LFT}(t_b)$, forcing all its successors to begin execution at their respective LSTs. As a result, these successors must be executed on the fastest available VMs to meet the workflow deadline $D_w$, thereby limiting opportunities for further slack minimization and energy reduction.

To address this limitation, we introduce the Level-based Deadline Distribution (LDD) approach \cite{DeadlineDistCite,DLD-Deadline}, which distributes the overall deadline proportionally across task levels. A \emph{level} is defined as a group of tasks that do not have dependencies among themselves. The level of a task $t_j$, denoted as $\lambda(t_j)$, is defined as follows: for a task with no predecessors, $\lambda(t_j) = 1$, and for all other tasks,

\begin{equation}
\label{eq:level}
\lambda(t_j) = \max_{t_k \in \operatorname{pred}(t_j)} \left( \lambda(t_k) + 1 \right)
\end{equation}

Let $\sigma(W) = \sum_{t_j \in W} wc_j$ represent the total worst-case computation time of all tasks in workflow $W$, and let $\sigma(\lambda(t_j))$ denote the cumulative worst-case computation time of tasks in level $\lambda(t_j)$. The level-based deadline assigned to task $t_j$, denoted as $\delta(t_j)$, is then given by:

\begin{equation}
\label{eq:level-dd}
\delta(t_j) = \delta(\lambda(t_j)-1) + (D_w - A_w) \cdot \frac{\sigma(\lambda(t_j))}{\sigma(W)}
\end{equation}

This formulation allocates a fraction of the total workflow deadline to each task based on the execution time distribution across levels. To preserve feasibility, we bound $\delta(t_j)$ within the range $[\operatorname{EFT}(t_j), \operatorname{LFT}(t_j)]$.

The following modifications distinguish this approach from the original LEF method:
\begin{enumerate}
\item The slack value is redefined as $\mathscr{S}'(t_j) = \delta(t_j) - \operatorname{Ft}(t_j)$. Consequently, the modified minimum compute-power–frequency product becomes $\texttt{MIN-CPF}'(t_j) = \frac{wc_j}{\delta(t_j) - \operatorname{St}(t_j)}$.
\item The task list $\mathcal{L}$ is ordered based on non-decreasing values of $\delta(t_j)$, i.e., tasks with earlier level-based deadlines are scheduled first.
\end{enumerate}

Since $\delta(t_j)$ is bounded above by $\operatorname{LFT}(t_j)$, bottleneck tasks do not push the successors to begin at their respective latest start times. This enables potential energy savings for successor tasks by allowing flexibility in their scheduling. The complete methodology is summarized in \autoref{alg:DD-SLWSFA}.

\begin{algorithm}[tb!]
\setcounter{AlgoLine}{0}
\caption{Level-based Deadline Distributed Approach (LDD)}
\label{alg:DD-SLWSFA}
\SetKwInOut{Input}{Input}
\SetKwInOut{Output}{Output}
\SetKw{KwInit}{Initialize}
\DontPrintSemicolon

\Input{$W = (\mathbf{T}, \mathbf{E})$, \( \text{VM}(C)\)}

\Output{Schedule \( \mathcal{S} \) for workflow \( W \) with \( \mathcal{S}(t_j) = (f_k^{\text{VM}_l}, St(t_j), Ft(t_j)), \: \forall t_j \in \mathbf{T}\)}

Calculate EST, EFT, LST, LFT of every task in $W$ using \autoref{eq:est}, \ref{eq:eft}, \ref{eq:lft}, \ref{eq:lst}.

$\delta(t_j) \gets$ Calculate the level based deadline of all tasks in $W$ using \autoref{eq:level-dd} and bound $\delta(t_j)$ within the range $[\operatorname{EFT}(t_j), \operatorname{LFT}(t_j)]$

\KwInit{$\mathcal{L} \gets \textnormal{tasks of } W \textnormal{ sorted in earliest level-based deadline}$}

\For{\textnormal{each task} \( t_j \) \textnormal{in list} $\mathcal{L}$}{

Calculate MIN-CPF ($t_j$) = $\frac{wc_j}{\delta(t_j) - \operatorname{St}(t_j)}$

$\mathcal{S}(t_j) \gets$ Generate energy aware schedule for task $t_j$ ensuring deadline, reliability using \autoref{alg:ret} ($W$, \( \text{VM}(C)\), EST, EFT, LST, LFT, MIN-CPF ($t_j$))

}

\Return \( \mathcal{S} \)

\end{algorithm}

\subsection{Comparing LEF and LDD}
\label{ss:compare}
 \begin{figure}
    \centering
    \includegraphics[scale=0.75]{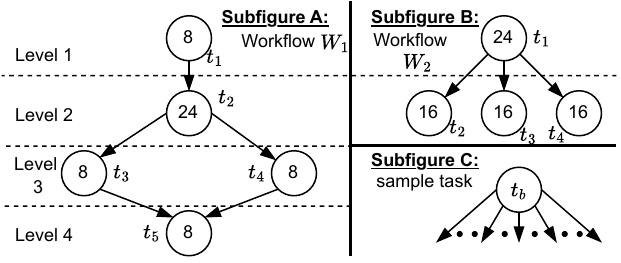}
    \caption{Subfigure A and B, are two workflows to compare LEF and LDD. Subfigure C shows a bottleneck task $t_b$}
    \label{fig:wf}
\end{figure}

Let us consider two workflows \( W_1 \) (shown in \autoref{fig:wf} A) and \( W_2 \) (shown in \autoref{fig:wf} B), and two virtual machines \( \text{VM}_1 \) and \( \text{VM}_2 \). The compute powers (CP) of \( \text{VM}_1 \) and \( \text{VM}_2 \) are 8 and 2, respectively. It means that the VMs can execute 8 MIPS when executed at maximum frequency. Both VMs can operate at two discrete frequencies, \( f_1 = 1 \) and \( f_2 = 0.5 \). The parameters for \( \text{VM}_1 \) are \( \alpha = 50 \) and \( \beta = 128 \), giving a critical frequency of approximately 0.579. For \( \text{VM}_2 \), \( \alpha = 40 \) and \( \beta = 64 \), yielding a critical frequency of approximately 0.678. The deadlines for \( W_1 \) and \( W_2 \) are 10 and 8 time units, respectively.

For \( W_1 \), the task order \( \mathcal{L} \) under the LEF approach is \( t_2, t_3, t_4, t_5, t_1 \). As \( t_3, t_4, t_5 \), and \( t_1 \) have identical worst-case execution times (in MIPS), their order can be permuted. In the LEF schedule, \( t_2 \), \( t_3 \), and \( t_4 \) are assigned to \( \text{VM}_1 \) at frequency 0.5, while \( t_5 \) and \( t_1 \) run at frequency 1. As a result: \( t_1 \) finishes at time $8/(1 * 8) = 1$, \( t_2 \) completes at $7 = (1 + 24/(0.5 * 8))$, \( t_3 \) and \( t_4 \) finish at $9 = (7+ 8/(0.5 * 8))$, \( t_5 \) concludes at $10 = (9 + 8/(1 * 8))$.
Thus, the deadline is met, and the total energy consumption is 1016 units.

In contrast, for LDD, the task order ($\mathcal{L}$) is \( t_1, t_2, t_3, t_4, t_5 \), where their maximum allowable finish times (\( \delta(t_j) \)) are:
$\delta(t_1) = 1.42,\: \delta(t_2) = 5.68,\: \delta(t_3) = 8.5,\: \delta(t_4) = 10,\: \delta(t_5) = 10$, calculated using \autoref{eq:level-dd}. The approach assigns \( t_1 \) and \( t_2 \) to \( \text{VM}_1 \) at frequency 1, and \( t_3, t_4, t_5 \) to \( \text{VM}_1 \) at frequency 0.5. This results in a total energy consumption of 1108 units. Hence, LEF performs better for \( W_1 \).

For \( W_2 \), LEF orders tasks in \( \mathcal{L} \) as \( t_1, t_2, t_3, t_4 \). Here, \( t_1 \) is scheduled on \( \text{VM}_1 \) at frequency $0.5$, and the remaining tasks at frequency $1$. Completion times are: \( t_1 \) completes at 6, \( t_2, t_3, t_4 \) complete at $10 = (6 + 4)$. This yields an energy cost of 1464 units.

In LDD approach for \( W_2 \), all tasks are assigned to \( \text{VM}_1 \), with \( t_1 \) at frequency $1$ and the others at $0.5$, resulting in a total energy consumption of 1326 units. Thus, LDD is preferable in this case.

From these examples, we observe that the LEF approach is generally more effective when the workflow contains tasks with fewer fan-outs, whereas LDD performs better in other scenarios. We use this observation to design of our adaptive approach, discussed in \autoref{ss:asmfr}.

 \subsection{\textnormal{Proposed Approach:} Adaptive Strategy Based on Maximum Fan-Out Ratio (ASMFR)}
 \label{ss:asmfr}

From \autoref{ss:compare}, we observe that the performance of the LDD and LEF approaches varies based on the structural characteristics of the workflow. Specifically, the LDD approach yields better results when a few nodes exhibit significantly higher fan-out compared to the rest, while LEF performs better in relatively balanced workflows. To quantify this, we introduce a parameter called the Maximum Fan-Out Ratio (MFR), defined as:
$\text{MFR} = \frac{d_{\max}}{n - 1}$, where \( d_{\max} \) represents the maximum fan-out of any node in the $W$, and \( n \) is the total number of tasks in the workflow \( W \).

Based on this observation, we propose a hybrid approach, ASMFR that selects between LEF and LDD dynamically. If the computed MFR is less than a predefined threshold \( Th \), the approach selects LEF; otherwise, it chooses LDD. This adaptive selection ensures that the most suitable scheduling heuristic is employed for the given workflow structure.

\begin{algorithm}[tb!]
\caption{Adaptive Strategy Based on Maximum Fan-Out Ratio (ASMFR)}
\label{alg:adaptive-mfr}
\footnotesize
\DontPrintSemicolon

\KwIn{$W = (\mathbf{T}, \mathbf{E})$, MFR, Threshold $Th$}
\KwOut{Chosen approach $\mathcal{A}$ for workflow $W$}

\eIf{$MFR < Th$}{
    $\mathcal{A} \gets$ \textbf{LEF} approach [\autoref{alg:LEF-SLWSA}] for $W$
}{
    $\mathcal{A} \gets$ \textbf{LDD} [\autoref{alg:DD-SLWSFA}] for  $W$
}

\Return $\mathcal{A}$

\end{algorithm}

 \subsection{Maintain Rolling Horizon by Dynamic Largest Energy Approach (Dy)}
 \label{subsec:Dynamic}
This approach dynamically adjusts the schedule $\mathbf{S}$ whenever a task completes earlier than its worst-case execution time, as observed in prior studies \cite{ExecTime}. Since tasks often finish sooner than the predicted upper bound, this behavior enables energy savings by turning off idle VMs and rescheduling the remaining tasks using the additional slack time.

The proposed strategy is event-driven: it dispatches tasks to VMs hosted on PMs and immediately updates the system upon task completion. Specifically, once a task completes, it is removed from the current schedule, and its actual finish time is used to recompute parameters for the remaining tasks in the workflow, in accordance with \autoref{eq:dag-const}.

To manage task readiness, we define a set called the \textit{Rolling Horizon}, denoted as $\mathcal{R}$, which contains all tasks eligible for dispatch. A task is considered ready if all its predecessors have completed execution. Initially, all tasks without predecessors are dispatched.

Upon completion of a task $t_j$, the cloud platform records its actual finish time and energy usage, removes $t_j$ from $\mathcal{R}$, and updates the state of the workflow $W$. Subsequently, the scheduling approach $\mathcal{A}$ (either LEF or LDD) is invoked to recompute the schedule by updating EST, EFT, LST, and LFT values based on the new state. The two components---dynamic adjustment (Dy) and the selected scheduling approach (LEF or LDD)---interact iteratively until all tasks in the workflow are executed.

\begin{algorithm}[tb!]
\setcounter{AlgoLine}{0}
\caption{Dynamic Largest Energy First Based Approach (Dy)}
\label{alg:dynamic}
\SetKwInOut{Input}{Input}
\SetKwInOut{Output}{Output}
\SetKw{KwInit}{Initialize}
\DontPrintSemicolon
\Input{$W = (\mathbf{T}, \mathbf{E})$, \( \text{VM}(C)\), Schedule \( \mathcal{S} \) for workflow \( W \) with $ \mathcal{S}(t_j) = (f_k^{\text{VM}_l}, St(t_j), Ft(t_j)) \forall j \in \mathbf{T}$ generated using LEF \autoref{alg:LEF-SLWSA} or LDD \autoref{alg:DD-SLWSFA}}
\Output{Pacing of task with no predecessor node on VM hosted on a PM and eventual task execution completion}

\KwInit{Rolling Horizon $\mathcal{R} \gets \emptyset$};

\For{\textnormal{each task} $t_j$ \textnormal{in} $W$} {
    \If{$pred(t_j) \cap W = \emptyset$ \textnormal{(\texttt{$t_j$ has no predecessor})}}{
         $\mathcal{R} \gets \mathcal{R}  \cup \{t_j\}$
         
         Dispatch task $t_j$ According to $\mathcal{S}$
    }
}

\While {$ \mathcal{R} \neq \emptyset$}{
    $t_j \gets$ First task to finish in $R$
    
    $\mathcal{R} \gets \mathcal{R} - \{t_j\} $ \texttt{(Remove $t_j$ from Rolling Horizon)}
    
    $W \gets W - \{t_j\} $ \texttt{(Modfify workflow $W$ by removing $t_j$)}
    
    Set EFT$(t_j)$ and LFT$(t_j)$ to $Ft_{\text{real}}(t_j)$ \texttt{(Since $t_j$ has already executed)}
    
    Update $Ft(t_j), E(t_j, .), R_j^{\textnormal{eff}}, R(W)$ and $E(W)$ based on $Ft_{\text{real}}(t_j)$ and VM mapping

}

\Return Placing $t_j$ on VM hosted on a PM
\end{algorithm}

\subsection{Worst Case Time Complexity Analysis}
We provide a very brief analysis of the worst-case time complexity of our proposed approaches in this section. As shown in \autoref{fig:scheduler-architecture}, ASMER chooses one of LEF and LDD based on Maximum Fan-out Raio (MFR) in $O(1)$ time. Max Fan-out can be calculated in $O(\mathbf{T})$ time as we need to scan all the tasks for number of successors. After it is chosen, the chosen approach (LEF/ LDD) and Dy operate alternative once for all the tasks in $W$. 

\subsubsection{Time Complexity Analysis of LEF or LDD}
Let $\mathbf{T}$ denote the number of tasks and $\mathbf{E}$ denote the edge set in workflow $W$ as discussed in \autoref{sec:task_model}.

\label{subsec:normal_tc}
\begin{itemize}
    \item We first implement Breadth First Search (BFS) algorithm, hence its time complexity is $\mathcal{O}(\mathbf{T} + \mathbf{E})$
    \item We then iterate through all VM types ($L$) present and find the frequency closest to \verb|MIN-CP|. Therefore, the time complexity of these steps is $\mathcal{O}(L \cdot \log{S_\text{l}})$, where $S_\text{l}$ is the total number of frequencies supported on a VM.
    \item Selection of minimum energy option from option list $\mathcal{O}$ takes $O(L)$ time in worst case.
\end{itemize}
Thus worst case time complexity of this part is $O(\mathbf{T}(\mathbf{T} + \mathbf{E} + L\cdot \log{S_\text{l}}+L))$, iterations are considered for all tasks in $W$.

\subsubsection{Time Complexity Analysis of Dy}
\label{subsec:dytc}
The time complexity of this approach is $O(\mathbf{T})$ for workflow $W$, where $\mathbf{T}$ denotes the total number of tasks in $W$. This complexity arises as each task without predecessors is dispatched individually, and relevant parameters such as finish time and energy consumption are updated for each task.

Hence overall time complexity of the proposed solution framework for workflow $W$ is $O(\mathbf{T}(\mathbf{T} + \mathbf{E} + L\cdot \log{S_\text{l}}+L) + \mathbf{T})$.

\section{Experiments and Results}
\label{subsec:Experiment}
In this section, we describe the experiments we performed to test the effectiveness of the proposed approaches. We developed a homegrown simulator for orchestration of workflow scheduling using C++  and Python similar to CloudSim \cite{cloudSim} and used benchmark workflows to generate results. 


\subsection{System Configurations}
We developed a simulation framework that allows us to customize the parameters for the system model, and evaluate the performance of our proposed approach. The parameter ranges used in our experiments are consistent with prior works such as \cite{Niyati,swain}. Note that frequency values are normalized to lie within the interval (0, 1) for each VM type, as discussed in \autoref{sec:system_model}.

In our experiments, we considered the following system parameters unless stated otherwise: the number of VM types $L$ ranged from 5 to 15; compute power $CP_l$ varied between 0.9 and 210; frequency levels $f_k^{\text{VM}_l}$ ranged from 0.12 GHz to 5.5 GHz; the number of frequency levels per VM type $S_l$ was between 4 and 6; the sensitivity coefficient $\psi$ was set between 3 and 7; and the base failure rate $r_0^l$ ranged from $10^{-6}$ to $10^{-4}$.

\subsection{Real life Workflow Benchmark}
\label{subsec:real-life-wf}
We used various scientific workflows available from Pegasus Workflow Management System \cite{PegasusWf} and other sources \cite{Scoop} like Montage, Sipht, Inspiral, Scoop etc,. Each workflow differs from others in terms of the DAG structure, maximum fan-out, etc, and provides a means to test our implementation. We have taken the deadline constraint $D_w$ of workflows by multiplying the time of execution of critical path when it is scheduled in VM with the highest computation power in maximum frequency possible by a deadline factor $df$ \cite{sota_deadline_Medara,Niyati}, ranging from 1 to 2.5. Similarly, Reliability constraint $R_w$ has been set randomly to a number $\in [0.9,1)$. These parameters are also varied accordingly to study their individual effect in our experiments. To incorporate the probabilistic nature of task execution times, we sampled the actual execution time $wc_j$ for each task from an exponential distribution \cite{expDist}. Specifically, we set the mean of the distribution to 75\% of the given data-dependent worst-case execution time and capped the maximum sampled value at $wc_j$.

\subsection{Approaches Considered for Performance Evaluation}
We evaluate our proposed approach against four state-of-the-art (SOTA) baselines. Two of these are established prior works, while the remaining two serve as idealized baselines representing optimal and aggressive execution strategies. The SOTA approaches considered are as follows:

\subsubsection{Reliability, rental cost, and Energy-aware Multi-workflow Scheduling with DVFS (REMSM-DVFS)} We compared our work with the approach proposed by Niar \textit{et al.} \cite{sota_deadline} referred to as REMSM-DVFS, which focuses on minimizing energy consumption and VM rental cost while improving reliability, modelled via a Weibull distribution in a multi-cloud environment. The method assigns level-based deadlines to tasks and schedules them using an earliest-deadline-first strategy. Each task is mapped to a VM using a score function that balances energy, cost, and reliability. Task replication is performed on energy-efficient VMs when needed, but DAG constraints are not enforced for replicas. Since our problem does not consider rental costs, we assume uniform rental pricing. Additionally, reliability is treated as a soft constraint, so $R_w$ may not always be satisfied.

\subsubsection{Reliability-Aware and Energy-Efficient Workflow Scheduling Algorithm (REWS)} 
We compared our work with Lingjuan \textit{et al.} \cite{sota_rel} approach referred to as REWS, which aims to minimize energy consumption while treating workflow reliability as a hard constraint. The approach introduces a sub-reliability prediction mechanism by decomposing the overall workflow reliability into individual reliability targets per task. For each task, the VM and operating frequency are selected to satisfy the sub-reliability constraint with minimal energy consumption. Since this approach does not consider workflow deadlines, we compare it against our methods under a relaxed constraint, assuming an infinite deadline ($D_w = \infty$).

\subsubsection{Static Optimal Solution (S-OPT)} We used \verb|Gurobi|, a solver widely recognized for handling constrained optimization problems \cite{gurobi_cite}. Due to the non-linear nature of the reliability model, the encoded problem falls under the class of Mixed-Integer Non-Linear Programming (MINLP) \cite{mit_nlp,gurobi_cite}. However, due to the high computational complexity, the solver failed to find optimal solutions for some workflows. S-OPT uses only static workflow information and does not use runtime decisions.
\subsubsection{Best Compute Power Approach (BCP)} It schedules all tasks on the VM with the highest compute power operating at the maximum available frequency. While this approach minimizes task completion times, it results in significantly high energy consumption. It does not consider runtime decisions. 

\subsection{Effect of different Workflows on Energy Consumption}

\begin{figure}[tb!]
    \centering
    \includegraphics[scale=0.58]{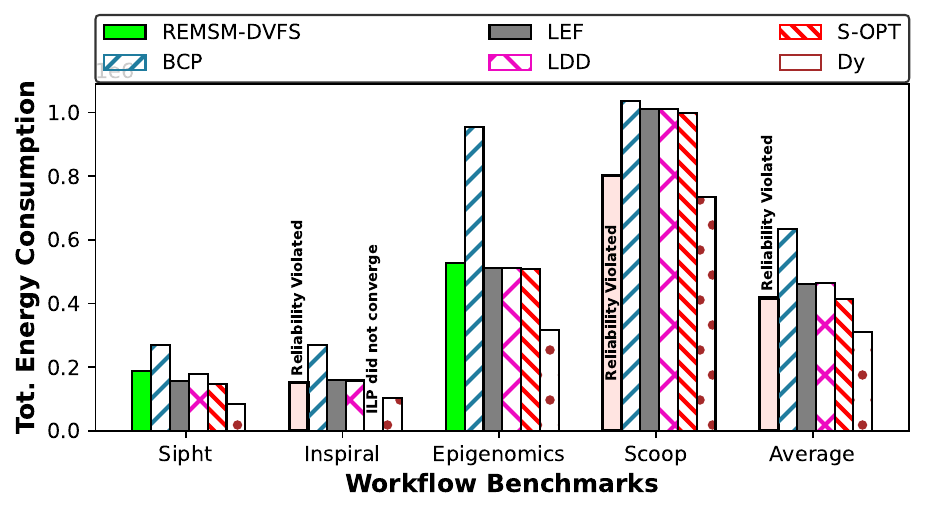}
    \caption{Comparison of total energy consumption across workflow benchmarks against REMSM-DVFS\cite{sota_deadline}. (deadline-based SOTA). Our proposed approach is best between LEF and LDD. \textcolor{customGreen}{\rule{1.5ex}{1.5ex}} denotes that $R_w$ is met, while \textcolor{customRed}{\rule{1.5ex}{1.5ex}} indicates otherwise. Our proposed approach is best between LEF and LDD. \texttt{Gurobi} solver did not converge for the Inspiral workflow.}
    \label{fig:ECVsW-SOTADeadline}
\end{figure}

\begin{figure}[!tb]
    \centering
    \includegraphics[scale=0.58]{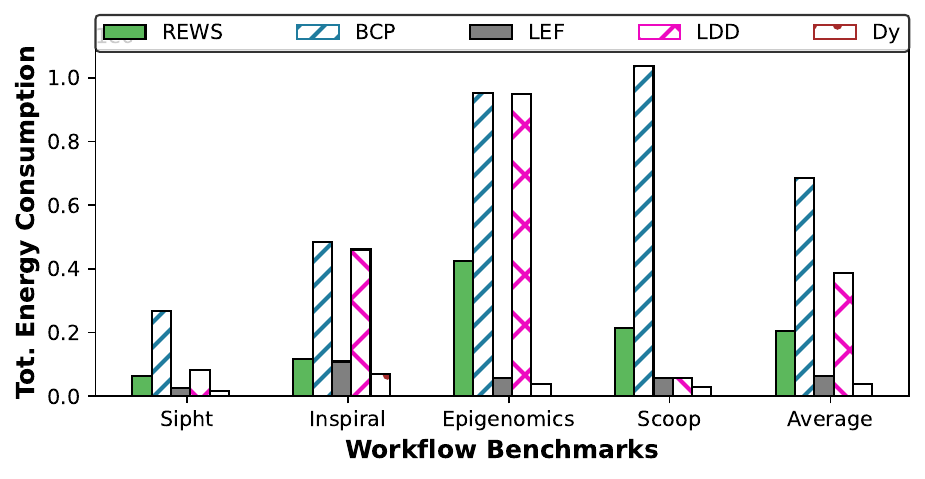}
    \caption{Comparison of total energy consumption across workflow benchmarks against REWS \cite{sota_rel} (non deadline-based SOTA). Our proposed approach is best between LEF and LDD.}
    \label{fig:ECVsW-SOTARel}
\end{figure}

To evaluate our approaches against deadline-based SOTA: REMSM-DVFS \cite{sota_deadline}, we set $df$ to 1.5 and used $R_w = 0.98$. The resulting energy consumption for various medium-sized workflow benchmarks is shown in \autoref{fig:ECVsW-SOTADeadline}. Our dynamic approach (Dy) (\autoref{alg:dynamic}) outperforms all other approaches ($\sim 40.3\%$), as it considers real-time task execution. In Sipht and Epigenomics, our proposed approach performs better ($\sim 7.7\%$) than the SOTA. Even though in Inspiral and Scoop, REMSM-DVFS seems to perform better, it does not adhere to reliability constraints as their objective function does not impose a hard constraint on reliability. In this particular value of deadline, LEF performs better than LDD because these workflows contain long straight chain of tasks for which slack minimization using LFT provides better results on average. We note that the LEF approach results in energy consumption closer to S-OPT (worse by $\sim 1.5\%$ on average), while the LDD approach performs worse by $\sim 3\%$ on average across the workflows analyzed in \autoref{fig:ECVsW-SOTADeadline}. Similar results follow with non deadline-based SOTA: REWS \cite{sota_rel} as shown in \autoref{fig:ECVsW-SOTARel}. Since the REWS approach focuses on meeting the reliability constraint, no violations of the constraint occur.

\subsection{Effect of number of tasks on Energy Consumption}

\begin{figure}[!tb]
    \includegraphics[scale=0.54]{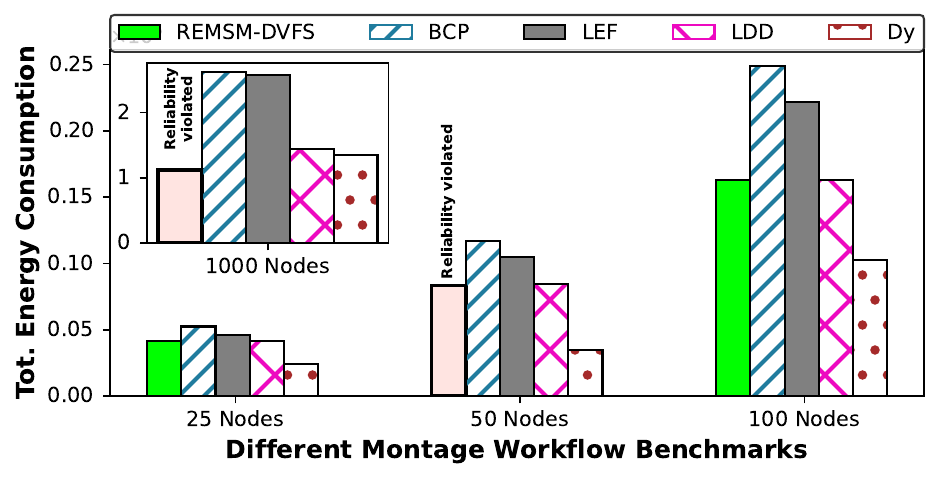}
    \caption{Comparison of energy consumption for the Montage workflow against REMSM-DVFS\cite{sota_deadline} (deadline-based SOTA), with varying numbers of nodes per task. \textcolor{customGreen}{\rule{1.5ex}{1.5ex}} denotes that $R_w$ is met, while \textcolor{customRed}{\rule{1.5ex}{1.5ex}} indicates otherwise. Our proposed approach is best between LEF and LDD.}
    \label{fig:ECvsMontage}
\end{figure}

We choose Montage workflow \cite{PegasusWf} to study the effect of a number of tasks on energy consumption with a tight $df = 1.2$. We can observe in \autoref{fig:ECvsMontage}, the intuitive trend of increase in energy with increase in number of tasks in the workflow for each algorithm. For 50 and 1000 node Montage workflow, REMSM-DVFS \cite{sota_deadline} fails to meet reliability constraint $R_w$. Our Dy approach achieves the lowest energy, and among static, BCP incurs the highest energy and lowest makespan in every case as it is energy-oblivious. The LDD approach performs better than the LEF approach because Montage workflow contains many tasks with multiple children and predecessors like the bottle-neck task in subfigure C of \autoref{fig:wf}.

\subsection{Effect of Deadline Factor $df$ on Energy Consumption}
\begin{figure}[!tb]
    \centering
    \includegraphics[scale=.54]{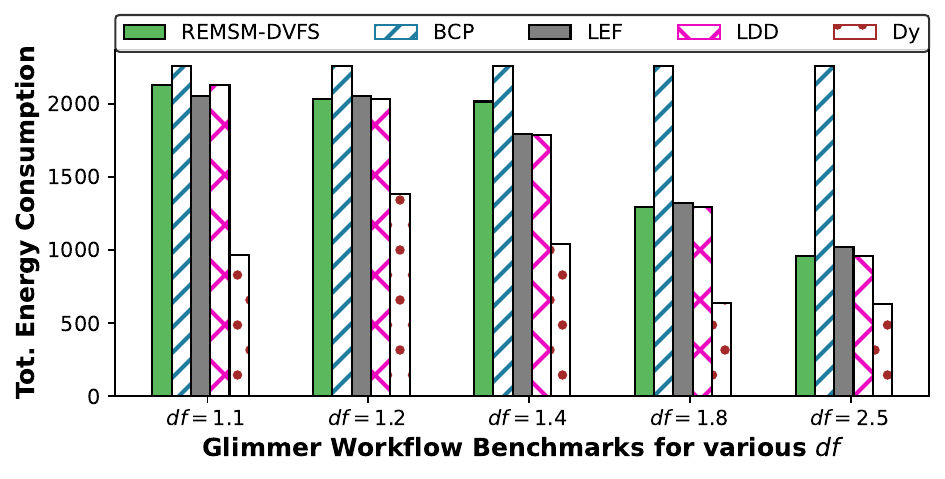}
    \caption{Comparison of energy consumption for Glimmer workflow \cite{Scoop}against REMSM-DVFS\cite{sota_deadline} (deadline-based SOTA) with $df$ varying from 1.1 (tight) to 2.5 (relaxed). Our proposed approach is best between LEF and LDD.}
    \label{fig:ECVsdf}
\end{figure}
To analyze the impact of the deadline factor ($df$) on energy consumption, we conducted experiments using Glimmer, a bio-informatics workflow for gene identification in microbial DNA \cite{Scoop}. We set $R_w$ to 0.99 and vary $df$ from 1.1 to 2.5, where lower $df$ values correspond to tighter deadlines. As shown in \autoref{fig:ECVsdf}, energy consumption generally decreases with increasing $df$. Notably, as BCP approach is deadline-oblivious, its energy consumption remains constant regardless of $df$. Dy approach consistently outperforms other algorithms due to its real-time execution consideration. For more relaxed $df$ values, REMSM-DVFS \cite{sota_deadline} and our proposed algorithms (excluding BCP) converge to similar energy consumption levels.
\subsection{Effect of Reliability Constraint $R_w$ on Energy Consumption}
\begin{figure}[!tb]
    \centering
    \includegraphics[scale=.52]{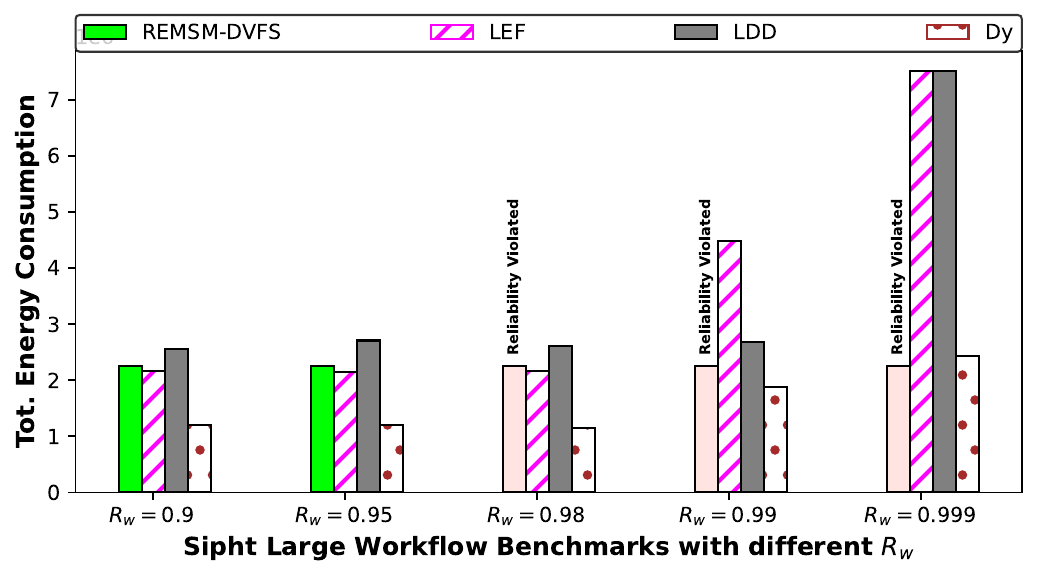}
    \caption{Comparison of energy consumption for Sipht workflow with 1000 nodes for REMSM-DVFS\cite{sota_deadline} (deadline-based SOTA). Our proposed approach is best between LEF and LDD.}
    \label{fig:ECVsR-SOTA-Deadline}

    \centering
    \includegraphics[scale=.57]{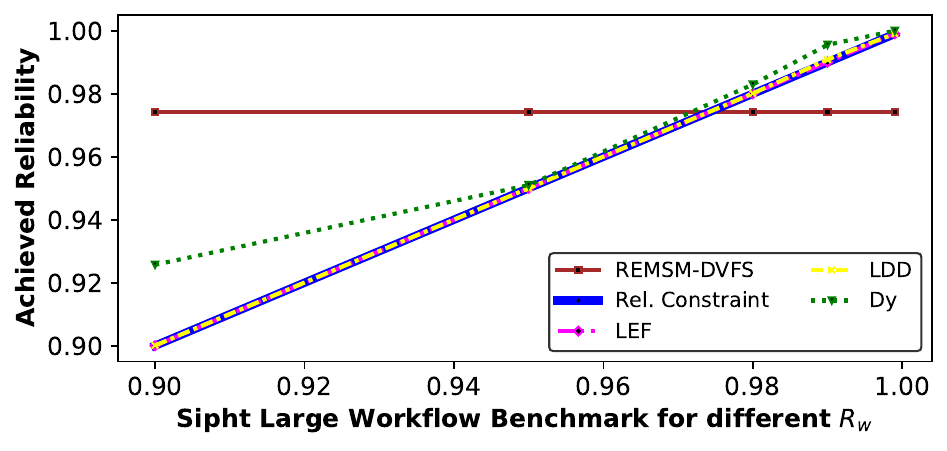}
    \caption{Achieved Reliability for different approaches as in \autoref{fig:ECVsR-SOTA-Deadline}. REMSM-DVFS violates reliability constraint for $R_w \geq 0.98$.}
    \label{fig:Rel-ECVsR-SOTA-Deadline}
\end{figure}

\begin{figure}[!tb]
    \centering
    \includegraphics[scale=.52]{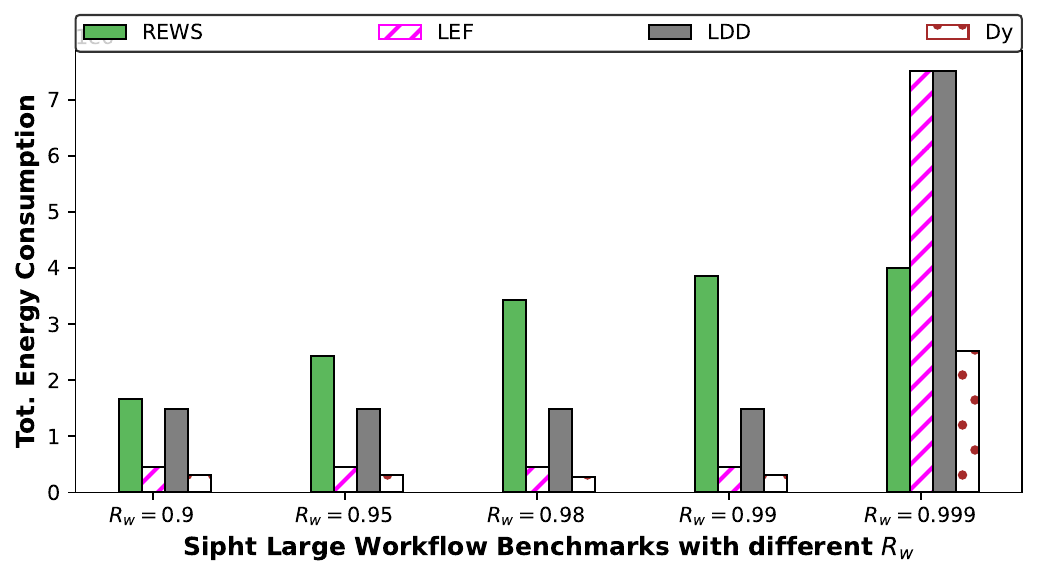}
    \caption{Comparison of energy consumption for Sipht Workflow with 1000 nodes for REWS\cite{sota_rel} (non deadline-based SOTA). Our proposed approach is best between LEF and LDD.}
    \label{fig:ECVsR-SOTA-Rel}

    \centering
    \includegraphics[scale=.57]{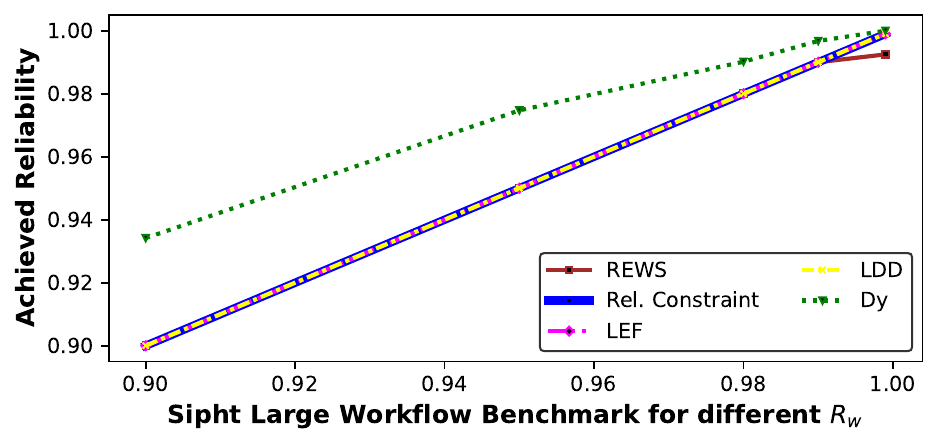}
    \caption{Achieved Reliability for different approaches as in \autoref{fig:ECVsR-SOTA-Rel}}
    \label{fig:Rel-ECVsR-SOTA-Rel}
\end{figure}

We next examine the effect of a strict reliability constraint on energy consumption using a large instance of the Sipht workflow with 1000 nodes. For this experiment, we set $df = 1.5$ and include only LEF, DD, and Dy approach to maintain clarity in the graph. Results are shown in \autoref{fig:ECVsR-SOTA-Deadline},\autoref{fig:Rel-ECVsR-SOTA-Deadline} for REMSM-DVFS\cite{sota_deadline} and \autoref{fig:ECVsR-SOTA-Rel},\autoref{fig:Rel-ECVsR-SOTA-Rel} for REWS\cite{sota_rel}. We observe that energy consumption rises with higher reliability constraint $R_w$, as tasks must execute at high frequencies to meet $R_w$, reducing opportunities for energy savings. Unlike our approaches, REMSM-DVFS \cite{sota_deadline}, focuses solely on maximizing reliability and hence fails to meet stricter constraints (\autoref{fig:Rel-ECVsR-SOTA-Deadline}) from $R_w=0.98$ to $R_w=0.999$, as evidenced by a stagnant reliability of 0.974 starting from $R_w=0.90$. Our proposed approaches consistently meet reliability constraints, with Dy approach yielding the best performance. As seen in \autoref{fig:ECVsR-SOTA-Rel}, our approaches outperform state-of-the-art (REWS) when $df = \infty$ while satisfying reliability constraints.

\section{Conclusion and Future work}
Through extensive experimentation, we concluded that dynamic scheduling could prove to be a reliable and robust means to schedule deadline-sensitive and reliability-critical multi-workflows. The cloud scheduling field is constantly evolving, and such new approaches are the need of the hour to keep up with the demand sustainably. Results show that our proposed approach outperforms many state-of-the-art solutions and is easy enough to adapt. 

Although our simulations make a probabilistic estimation of the execution time of tasks, in future works, task-based profiling could be targeted to estimate the execution time of tasks. Since the algorithm makes a real-time decision, working on the implementation and reducing time complexity is an important future task. Our work can also be extended to a multi-cloud environment by including bandwidth between different clouds, which can be presented in a future article.

\newpage
\bibliographystyle{IEEEtran}
\bibliography{main}

\begin{thebibliography}{10}
\providecommand{\url}[1]{#1}
\csname url@samestyle\endcsname
\providecommand{\newblock}{\relax}
\providecommand{\bibinfo}[2]{#2}
\providecommand{\BIBentrySTDinterwordspacing}{\spaceskip=0pt\relax}
\providecommand{\BIBentryALTinterwordstretchfactor}{4}
\providecommand{\BIBentryALTinterwordspacing}{\spaceskip=\fontdimen2\font plus
\BIBentryALTinterwordstretchfactor\fontdimen3\font minus \fontdimen4\font\relax}
\providecommand{\BIBforeignlanguage}[2]{{%
\expandafter\ifx\csname l@#1\endcsname\relax
\typeout{** WARNING: IEEEtran.bst: No hyphenation pattern has been}%
\typeout{** loaded for the language `#1'. Using the pattern for}%
\typeout{** the default language instead.}%
\else
\language=\csname l@#1\endcsname
\fi
#2}}
\providecommand{\BIBdecl}{\relax}
\BIBdecl

\bibitem{Power_model}
M.~Dayarathna, Y.~Wen, and R.~Fan, ``Data center energy consumption modeling: A survey,'' \emph{IEEE Communications Surveys and Tutorials}, 2016.

\bibitem{sota_rel}
L.~Ye, Y.~Xia, S.~Tao, C.~Yan, R.~Gao, and Y.~Zhan, ``Reliability-aware and energy-efficient workflow scheduling in iaas clouds,'' \emph{IEEE Transactions on Automation Science and Engineering}, 2023.

\bibitem{trial}
M.~Iverson, F.~Ozguner, and L.~Potter, ``Statistical prediction of task execution times through analytic benchmarking for scheduling in a heterogeneous environment,'' \emph{IEEE Transactions on Computers}, vol.~48, no.~12, pp. 1374--1379, 1999.

\bibitem{Roll-H}
X.~Zhu, H.~Chen, L.~T. Yang, and S.~Yin, ``Energy-aware rolling-horizon scheduling for real-time tasks in virtualized cloud data centers,'' in \emph{2013 IEEE 10th International Conference on High Performance Computing and Communications and 2013 IEEE International Conference on Embedded and Ubiquitous Computing}, 2013.

\bibitem{pbmodel}
P.~Chundi, R.~Narasimhan, D.~J. Rosenkrantz, and S.~S. Ravi, ``Active client primary-backup protocols,'' in \emph{Proceedings of the Fourteenth Annual ACM Symposium on Principles of Distributed Computing}, ser. PODC '95.\hskip 1em plus 0.5em minus 0.4em\relax Association for Computing Machinery, 1995, p. 264.

\bibitem{ECao}
M.~C. E~Cao, Saira~Musa, ``Energy and reliability-aware task scheduling for cost optimization of dvfs-enabled cloud workflows,'' \emph{IEEE TRANSACTIONS ON CLOUD COMPUTING}, 2023.

\bibitem{sota_deadline}
A.~Taghinezhad-Niar and J.~Taheri, ``Reliability, rental-cost and energy-aware multi-workflow scheduling on multi-cloud systems,'' \emph{IEEE Transactions on Cloud Computing}, 2023.

\bibitem{Niyati}
M.~Ghose, K.~P. Pandey, N.~Chaudhari, and A.~Sahu, ``Soft reliability aware scheduling of real-time applications on cloud with mttf constraints,'' in \emph{2023 IEEE/ACM 23rd International Symposium on Cluster, Cloud and Internet Computing (CCGrid)}, 2023.

\bibitem{swain}
C.~K. Swain and A.~Sahu, ``Reliability-ensured efficient scheduling with replication in cloud environment,'' \emph{IEEE Systems Journal}, 2022.

\bibitem{sota_deadline_Mousavi}
S.~S. Mousavi~Nik, M.~Naghibzadeh, and Y.~Sedaghat, ``Task replication to improve the reliability of running workflows on the cloud,'' \emph{Cluster Computing}, vol.~24, 2021.

\bibitem{Realtime}
X.~Ma, H.~Xu, H.~Gao, and M.~Bian, ``Real-time multiple-workflow scheduling in cloud environments,'' \emph{IEEE Transactions on Network and Service Management}, vol.~18, no.~4, pp. 4002--4018, 2021.

\bibitem{pc6}
B.~Akesson, M.~Nasri, G.~Nelissen, S.~Altmeyer, and R.~I. Davis, ``An empirical survey-based study into industry practice in real-time systems,'' in \emph{2020 IEEE Real-Time Systems Symposium (RTSS)}, 2020, pp. 3--11.

\bibitem{pc7}
M.~Nasri and B.~B. Brandenburg, ``An exact and sustainable analysis of non-preemptive scheduling,'' in \emph{2017 IEEE Real-Time Systems Symposium (RTSS)}, 2017, pp. 12--23.

\bibitem{pc8}
B.~Yalcinkaya, M.~Nasri, and B.~B. Brandenburg, ``An exact schedulability test for non-preemptive self-suspending real-time tasks,'' in \emph{2019 Design, Automation \& Test in Europe Conference \& Exhibition (DATE)}, 2019, pp. 1228--1233.

\bibitem{pc9}
M.~Nasri, T.~Chantem, G.~Bloom, and R.~M. Gerdes, ``On the pitfalls and vulnerabilities of schedule randomization against schedule-based attacks,'' in \emph{2019 IEEE Real-Time and Embedded Technology and Applications Symposium (RTAS)}, 2019, pp. 103--116.

\bibitem{sota_deadline_Medara}
R.~Medara and R.~Singh, ``Energy efficient and reliability aware workflow task scheduling in cloud environment,'' \emph{Wireless Personal Communications}, 2021.

\bibitem{On_Line}
S.~Liu, G.~Quan, and S.~Ren, ``On-line scheduling of real-time services for cloud computing,'' in \emph{2010 6th World Congress on Services}, 2010, pp. 459--464.

\bibitem{VM-Model}
G.~L. Stavrinides and H.~D. Karatza, ``An energy-efficient, qos-aware and cost-effective scheduling approach for real-time workflow applications in cloud computing systems utilizing dvfs and approximate computations,'' \emph{Future Gener. Comput. Syst.}, 2019.

\bibitem{Inf_PM}
T.~Erl, R.~Puttini, and Z.~Mahmood, \emph{Cloud Computing: Concepts, Technology \& Architecture}, 1st~ed.\hskip 1em plus 0.5em minus 0.4em\relax USA: Prentice Hall Press, 2013.

\bibitem{InfPM2}
G.~Brataas, N.~Herbst, S.~Ivansek, and J.~Polutnik, ``Scalability analysis of cloud software services,'' in \emph{2017 IEEE International Conference on Autonomic Computing (ICAC)}, 2017.

\bibitem{wcCitation}
J.~Bin, S.~Girbal, D.~Gracia~Pérez, A.~Grasset, and A.~Merigot, ``Studying co-running avionic real-time applications on multi-core cots architectures,'' 2014, pp. 35--40.

\bibitem{TransFail-Cite}
G.~Xie, G.~Zeng, Y.~Chen, Y.~Bai, Z.~Zhou, R.~Li, and K.~Li, ``Minimizing redundancy to satisfy reliability requirement for a parallel application on heterogeneous service-oriented systems,'' \emph{IEEE Transactions on Services Computing}, 2020.

\bibitem{ReliabilityCite}
D.~Zhu, R.~Melhem, and D.~Mosse, ``The effects of energy management on reliability in real-time embedded systems,'' 2004.

\bibitem{PB_Cite}
G.~Yao, X.~Li, Q.~Ren, and R.~Ruiz, ``Failure-aware elastic cloud workflow scheduling,'' \emph{IEEE Transactions on Services Computing}, 2023.

\bibitem{pc1}
E.~Sisinni, A.~Saifullah, S.~Han, U.~Jennehag, and M.~Gidlund, ``Industrial internet of things: Challenges, opportunities, and directions,'' \emph{IEEE Transactions on Industrial Informatics}, vol.~14, no.~11, pp. 4724--4734, 2018.

\bibitem{pc2}
J.~Pan, R.~Jain, S.~Paul, T.~Vu, A.~Saifullah, and M.~Sha, ``An internet of things framework for smart energy in buildings: Designs, prototype, and experiments,'' \emph{IEEE Internet of Things Journal}, vol.~2, no.~6, pp. 527--537, 2015.

\bibitem{pc3}
A.~Saifullah, D.~Ferry, J.~Li, K.~Agrawal, C.~Lu, and C.~D. Gill, ``Parallel real-time scheduling of dags,'' \emph{IEEE Transactions on Parallel and Distributed Systems}, vol.~25, no.~12, pp. 3242--3252, 2014.

\bibitem{pc4}
\BIBentryALTinterwordspacing
A.~Bhuiyan, Z.~Guo, A.~Saifullah, N.~Guan, and H.~Xiong, ``Energy-efficient real-time scheduling of dag tasks,'' \emph{ACM Trans. Embed. Comput. Syst.}, vol.~17, no.~5, Sep. 2018. [Online]. Available: \url{https://doi.org/10.1145/3241049}
\BIBentrySTDinterwordspacing

\bibitem{pc5}
Z.~Guo, A.~Bhuiyan, D.~Liu, A.~Khan, A.~Saifullah, and N.~Guan, ``Energy-efficient real-time scheduling of dags on clustered multi-core platforms,'' in \emph{2019 IEEE Real-Time and Embedded Technology and Applications Symposium (RTAS)}, 2019, pp. 156--168.

\bibitem{LSTEFT}
Q.~Wu, F.~Ishikawa, Q.~Zhu, Y.~Xia, and J.~Wen, ``Deadline-constrained cost optimization approaches for workflow scheduling in clouds,'' \emph{IEEE Transactions on Parallel and Distributed Systems}, 2017.

\bibitem{DeadlineDistCite}
A.~Taghinezhad-Niar, S.~Pashazadeh, and J.~Taheri, ``Energy-efficient workflow scheduling with budget-deadline constraints for cloud,'' \emph{Computing}, vol. 104, 2022.

\bibitem{DLD-Deadline}
V.~Arabnejad, K.~Bubendorfer, and B.~Ng, ``Deadline distribution strategies for scientific workflow scheduling in commercial clouds,'' in \emph{2016 IEEE/ACM 9th International Conference on Utility and Cloud Computing (UCC)}, 2016.

\bibitem{ExecTime}
M.~H.~J. Saldanha and A.~K. Suzuki, ``Determining the probability distribution of execution times,'' in \emph{2021 IEEE Symposium on Computers and Communications (ISCC)}, 2021.

\bibitem{cloudSim}
R.~N. Calheiros, R.~Ranjan, A.~Beloglazov, C.~A. F.~D. Rose, and R.~Buyya, ``Cloudsim: a toolkit for modeling and simulation of cloud computing environments and evaluation of resource provisioning algorithms.'' \emph{Softw., Pract. Exper.}, vol.~41, no.~1, pp. 23--50, 2011.

\bibitem{PegasusWf}
G.~Juve, A.~Chervenak, E.~Deelman, S.~Bharathi, G.~Mehta, and K.~Vahi, ``Characterizing and profiling scientific workflows,'' \emph{Future Gener. Comput. Syst.}, 2013.

\bibitem{Scoop}
L.~Ramakrishnan and B.~Plale, ``A multi-dimensional classification model for scientific workflow characteristics,'' in \emph{Proceedings of the 1st International Workshop on Workflow Approaches to New Data-Centric Science}, ser. Wands '10.\hskip 1em plus 0.5em minus 0.4em\relax New York, NY, USA: Association for Computing Machinery, 2010.

\bibitem{expDist}
M.~H.~J. Saldanha and A.~K. Suzuki, ``Determining the probability distribution of execution times,'' in \emph{2021 IEEE Symposium on Computers and Communications (ISCC)}, 2021.

\bibitem{gurobi_cite}
\BIBentryALTinterwordspacing
T.~Achterberg, ``Gurobi optimization releases new, groundbreaking version of its industry-leading mathematical programming solver,'' Mar. 2020, introduced non-convex quadratic optimization and MINLP capabilities. [Online]. Available: \url{https://shorturl.at/H0aLR}
\BIBentrySTDinterwordspacing

\bibitem{mit_nlp}
\BIBentryALTinterwordspacing
M.~O.~R. Center, ``Nonlinear programming,'' Web Resource, 2025. [Online]. Available: \url{https://web.mit.edu/15.053/www/AMP-Chapter-13.pdf}
\BIBentrySTDinterwordspacing

\end{thebibliography}
\end{document}